\documentclass[11pt]{article}
\usepackage{Blois,epsfig}

\def\be{\begin{equation}}
\def\ee{\end{equation}}
\def\bea{\begin{eqnarray}}
\def\eea{\end{eqnarray}}

\begin{document}
\vspace*{2cm}
\begin{center}
\Large{\textbf{XIth International Conference on\\ Elastic and Diffractive Scattering\\ Ch\^{a}teau de Blois, France, May 15 - 20, 2005}}
\end{center}

\vspace*{2cm}
\title{Medium-Modified Jet Shapes in Heavy-Ion Collisions}

\author{ C.A. SALGADO }

\address{Department of Physics, CERN, Theory Division\\
CH-1211 Geneva (Switzerland)}

\maketitle\abstracts{Jet quenching has been established as one of the main 
tools to study the properties of the medium produced in heavy ion collisions.
Most of the experimental effort has been, up to now, on the measurements of
inclusive particle suppression. This observable suffers, however, of
limitations due to different trigger-bias effects. The study of jets (or
particle correlations) in a medium is the most promising way out for a better
characterization of the medium properties. I will present how these more
differential measurements can be used to study not only the density of the
medium (the traditional parameter fixed by jet quenching measurements) but
also more dynamical quantities as flow fields.}

One of the more striking observations at RHIC so far is the strong suppression
of the yields of particles produced at high transverse momentum in central
nucleus-nucleus collisions \cite{Adcox:2001jp}. The most
widely accepted interpretation of this phenomenon is in terms of energy loss
due to medium-induced gluon radiation \cite{reviews}. 
The peculiar angular structure of this
radiation, affected by formation time effects through LPM suppression,
predicts a broadening of the jet-like signals associated to the high-$p_t$
particles when compared with the corresponding ones in proton-proton
collisions \cite{Baier:1996sk,Salgado:2003rv}. 

\begin{figure}
\begin{minipage}{0.5\textwidth}
\includegraphics[width=1\textwidth]{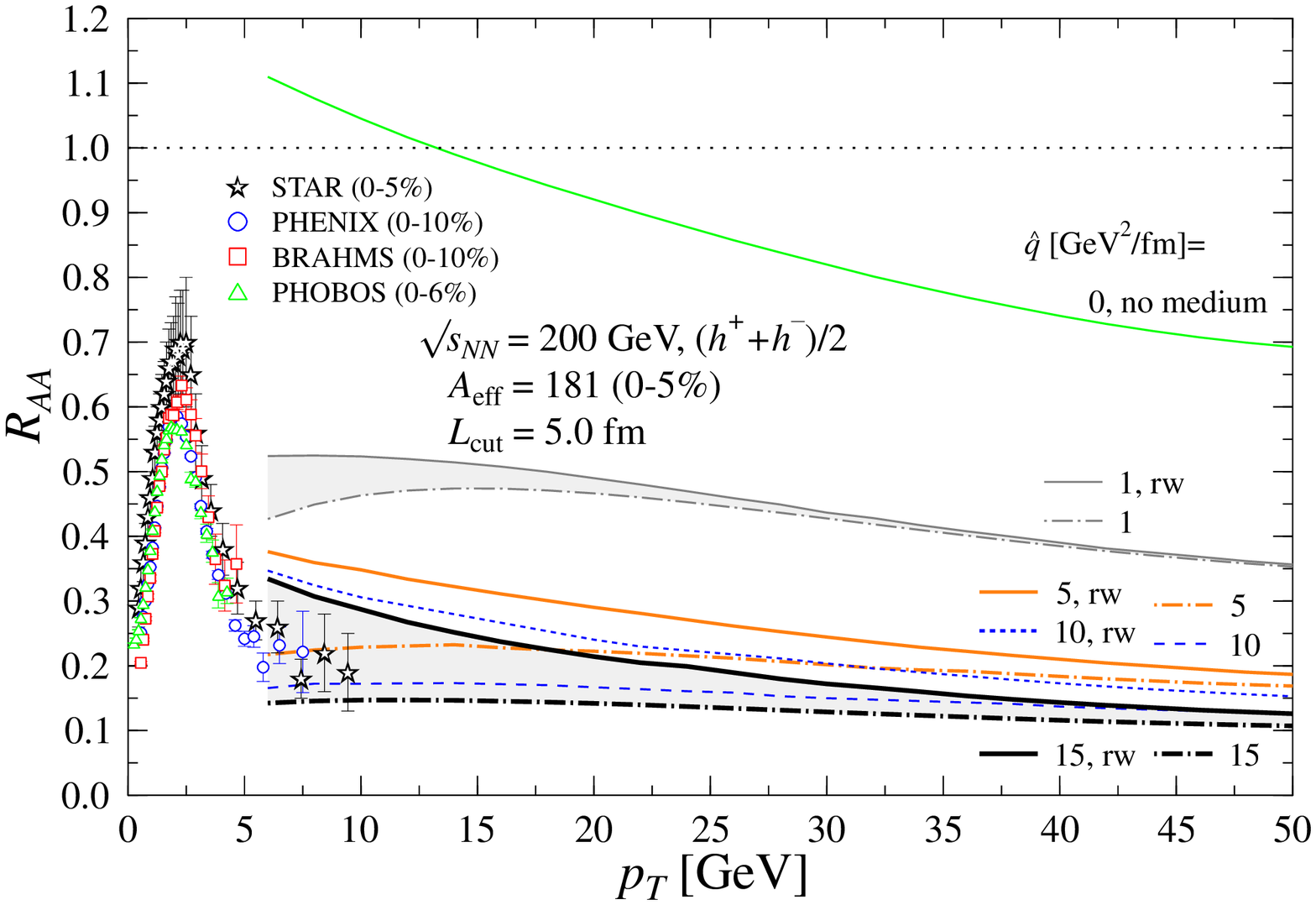}
\end{minipage}
\hfill
\begin{minipage}{0.5\textwidth}
\includegraphics[height=1\textwidth,angle=90]{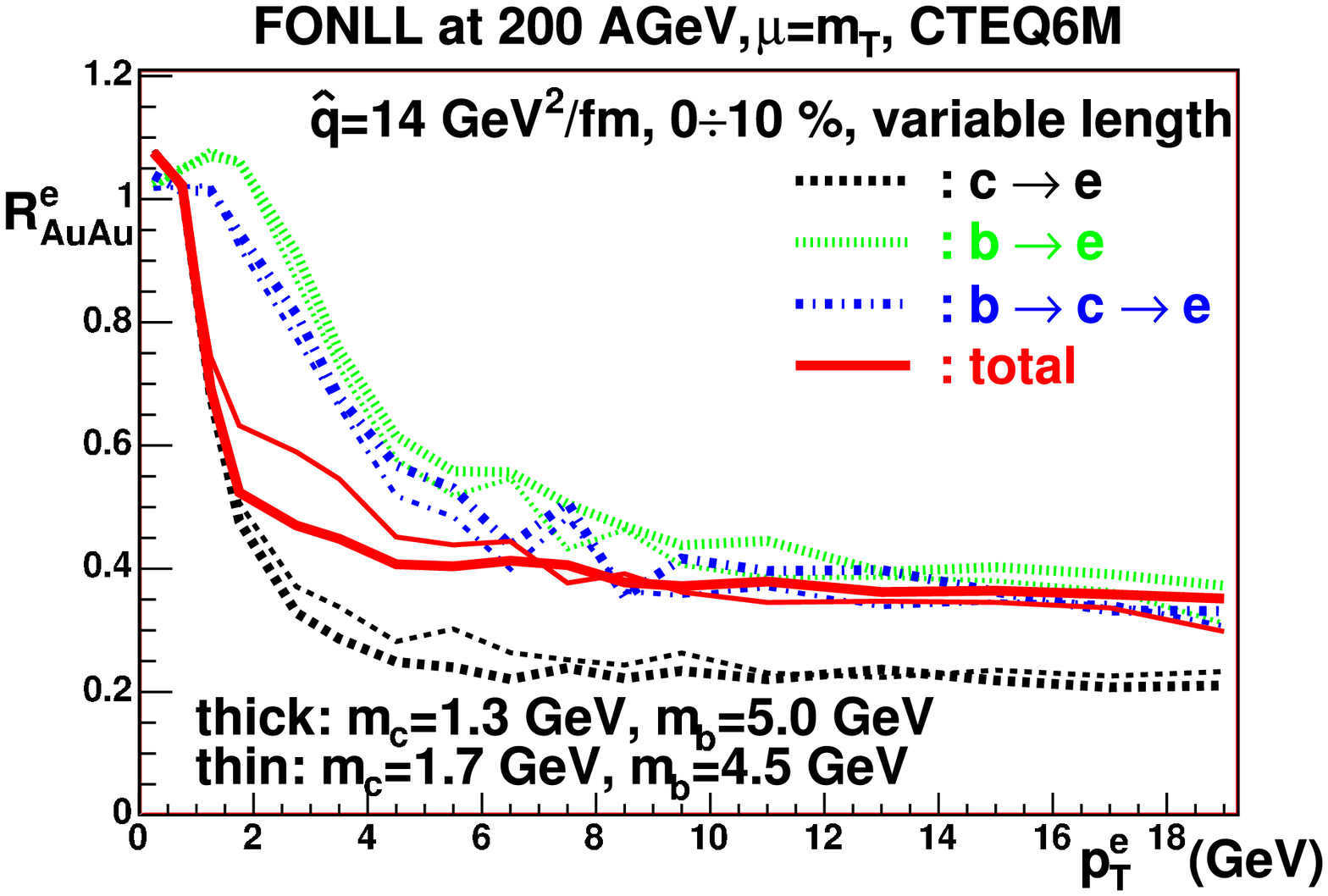}
\end{minipage}
\caption{Left:
nuclear modification factor $R_{AA}$ for charged particles in central AuAu
collisions at $\sqrt{s}=200$ GeV compared with theoretical curves
~\protect\cite{Eskola:2004cr} for different values of $\hat q$. Right:
Prediction ~\protect\cite{noshq}
for the suppression of electrons coming from the decay of charm
quarks in central AuAu collisions at $\sqrt{s}=200$ GeV.}
\label{fig1}
\end{figure}

{\it Inclusive particles}. The evolution of a high-$p_t$ parton shower
is affected by the medium created in heavy ion collisions. At the leading
particle level, the additional energy loss translates into a suppression of
the high-$p_t$ yields due to the steeply falling perturbative spectrum. The
only free parameter describing the medium-induced gluon radiation is the
transport coefficient $\hat q$ with the meaning of the transverse momentum
transferred to the emitted gluon per mean free path. This energy loss is normally
included \cite{Wang:1996yh}
as a medium-modification of the fragmentation functions, through a
convolution of the vacuum fragmentation function and the probability of an
additional medium-induced energy loss $P(\epsilon,\hat q, L)$. The main goal
in this type of analysis is to obtain the best value of the transport
coefficient which fits the experimental data. All the medium properties 
accessible by this probe are
encoded into this single variable. In Fig. \ref{fig1} we plot the experimental
data on light-meson suppression observed at RHIC together with the model
calculations using different values of $\hat q$. The data is well reproduced
for $\hat q=$ 5...15 GeV$^2$/fm. Two comments are in order here. On the one
hand the large uncertainty in the determination of this value is an intrinsic
limitation of inclusive particle suppression as a measurement when the studied
medium is very dense. The origin is a trigger bias effect that selects only
those particles produced close to the surface due to the steeply falling
spectrum of perturbatively produced partons \cite{Eskola:2004cr,Muller:2002fa}. 
On the other hand, this value is
more than five times larger than estimates based on perturbative coupling
of the traversing particles with the medium \cite{Baier:2002tc,Eskola:2004cr}. 
This has been interpreted as a
signal of strong non-perturbative effects on the medium 
\cite{Eskola:2004cr} or the coupling of the
jet to dynamical properties of the medium as a flow field 
\cite{Armesto:2004pt}. Both explanations
open new possibilities for the study of high-$p_t$ particles as probes to
characterize the medium.
 
One possibility for further constrain the value of $\hat q $ is by changing
the identity of the parent parton. In general color and mass factors imply a
larger energy loss for gluons than for quarks and also larger for light than
for heavy quarks \cite{Dokshitzer:2001zm}. 
The use of this property has been explicitly worked-out in
\cite{Armesto:2005iq} for both RHIC and the LHC. 
 At present, heavy meson identification has not
been possible at RHIC and the only information comes from the weak decays into
electrons. In Fig. \ref{fig1} our prediction for the electrons coming from 
the decays of $D$ and $B$ mesons is presented
using the same value of $\hat q$ as obtained in the fit to the light mesons.
Preliminary data \cite{qm05} shows a strong suppression, compatible with a 
dominant contribution of the charm quark to the observed electrons spectrum.
More work to understand the relative normalization of $c$ and $b$ production at RHIC energies is needed for a better understanding of the dynamical origin of 
this suppression.

{\it Medium-modification of the jet shapes}. The structure of the jets is
expected to be strongly modified when developed
in a medium. The larger emission angle of the medium-induced spectrum
translates into a broadening of the jet shapes. A main issue in jet
reconstruction (even
in more elementary collisions) is the energy calibration. The amount of
energy deposited into a jet cone of radius
$R\equiv\sqrt{\Delta\phi^2+\Delta\eta^2}$ is
affected by two opposite effects,  namely the
out-of-cone fluctuations (which increase with the cone radius) and the
background fluctuations (which decrease with the cone radius).
Dealing with the  high-multiplicity background
is the main difficulty in jet studies in heavy ion
collisions. From a theoretical point of view, it is then, essential to
identify jet observables in which the medium modification is not largely
affected by the background.

\begin{figure}
\begin{minipage}{0.5\textwidth}
\begin{center}
\includegraphics[width=0.8\textwidth]{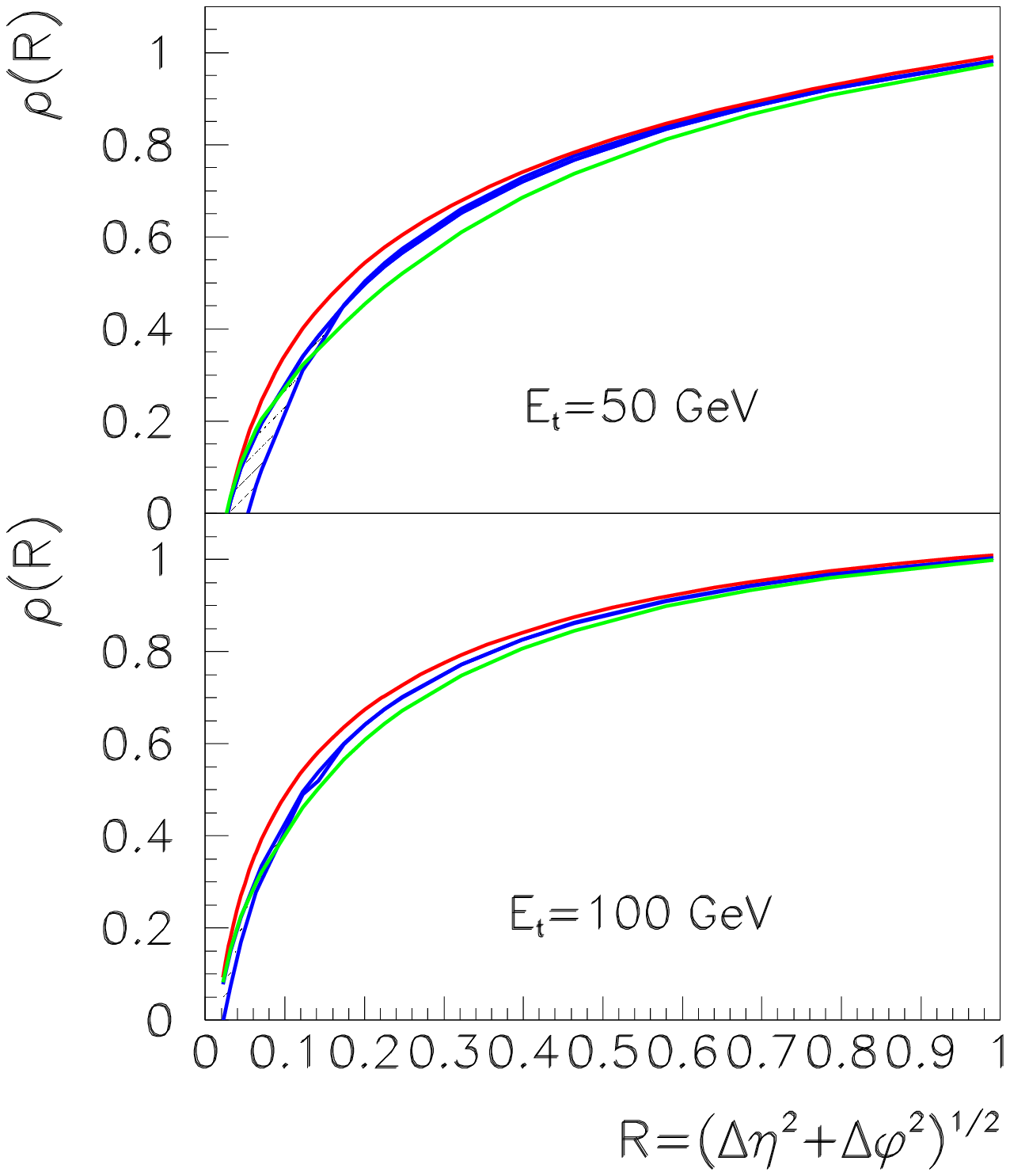}
\end{center}
\end{minipage}
\hfill
\begin{minipage}{0.5\textwidth}
\begin{center}
\includegraphics[width=0.8\textwidth]{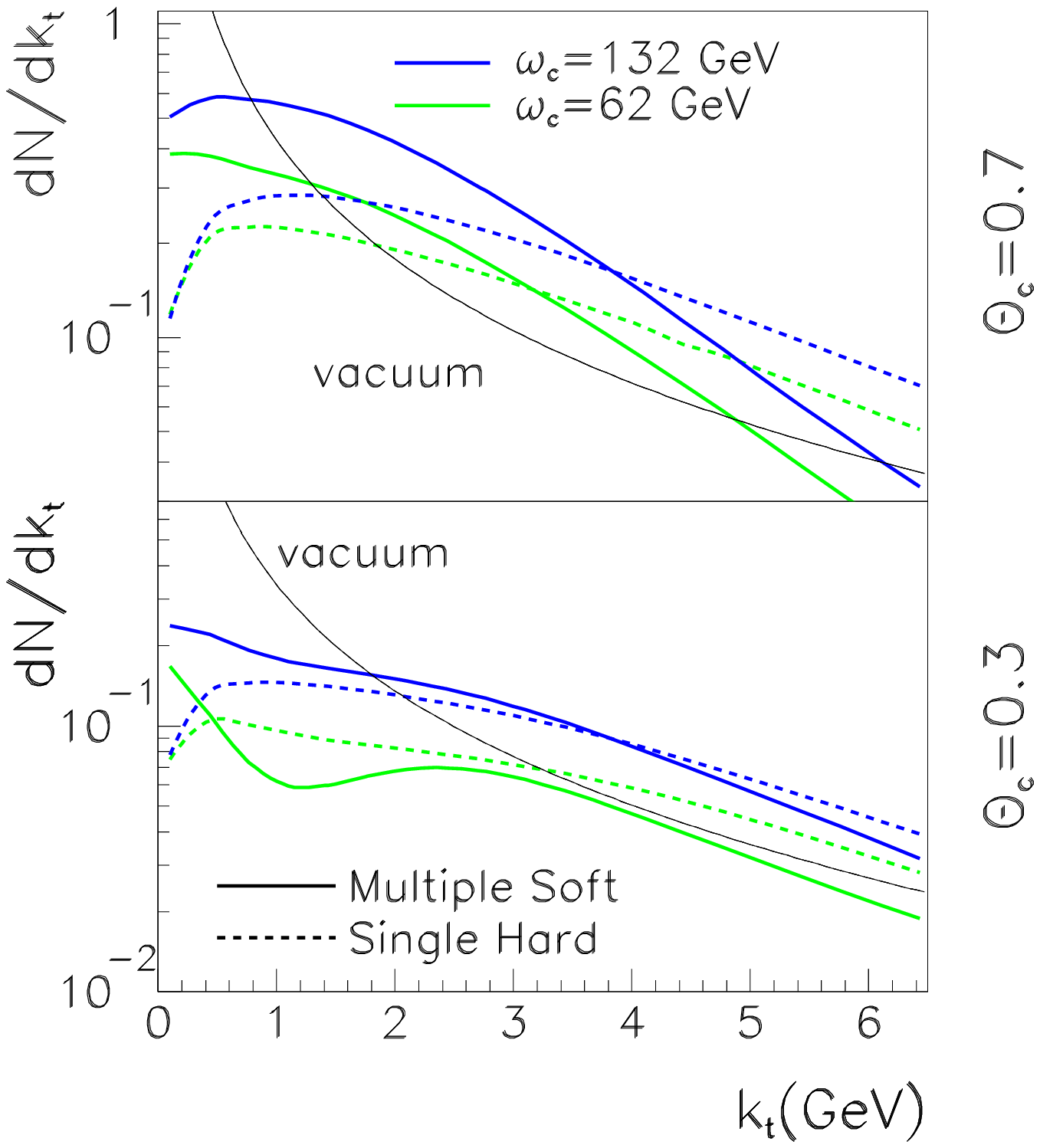}
\end{center}
\end{minipage}
\caption{Left: Fraction of the jet energy inside a cone
$R=\sqrt{\Delta \eta^2+\Delta\phi^2}$ for a 50 GeV and 100 GeV quark jet
fragmenting in the vacuum (red curves) and a hot medium. Right: Comparison of
the vacuum and medium-induced part of the gluon multiplicity distributions
inside a cone jet of size $R=\Theta_c$. Figures taken from
~\protect\cite{Salgado:2003rv}.}
\label{figjet}
\end{figure}

In \cite{Salgado:2003rv}
the first study of this type of observables was performed. In
particular we found a small broadening in the energy distributions inside a
jet when computing the fraction $\rho(R)$ of the total jet energy deposited
within a subcone of radius $R$. Physically, the gluons emitted at larger
angles
are softer and unable to redistribute a sizable amount of energy. As a
consequence,
the additional out-of-cone fluctuations are not enhanced dramatically.
 This general result is
independent on imposing small-$p_t$ cuts to the observed associated radiation,
see Fig. \ref{figjet}
If the jet energy distribution is not
modified by medium-effects the different structure of the radiation should
manifest in the multiplicity distributions. In Fig. \ref{figjet} we plot the
additional number of medium-induced gluons as a function of their transverse
momentum with respect to the jet axis for different values of the cone-radius
$R=\Theta_c$.
By imposing different cuts to the energy spectrum the sensitivity of this
observable to background subtraction is shown to be small.
 It is worth noting that even though the present calculations
lack of several physical mechanisms as hadronization etc,
the main conclusions are
independent on the actual realization of the model and depend solely on the
general properties of the medium-induced gluon radiation; in particular the
non-divergency of the spectrum in the infrared and collinear limits. These
properties are given by formation time effects and kinematics.

\begin{figure}
\begin{minipage}{0.52\textwidth}
\includegraphics[width=\textwidth]{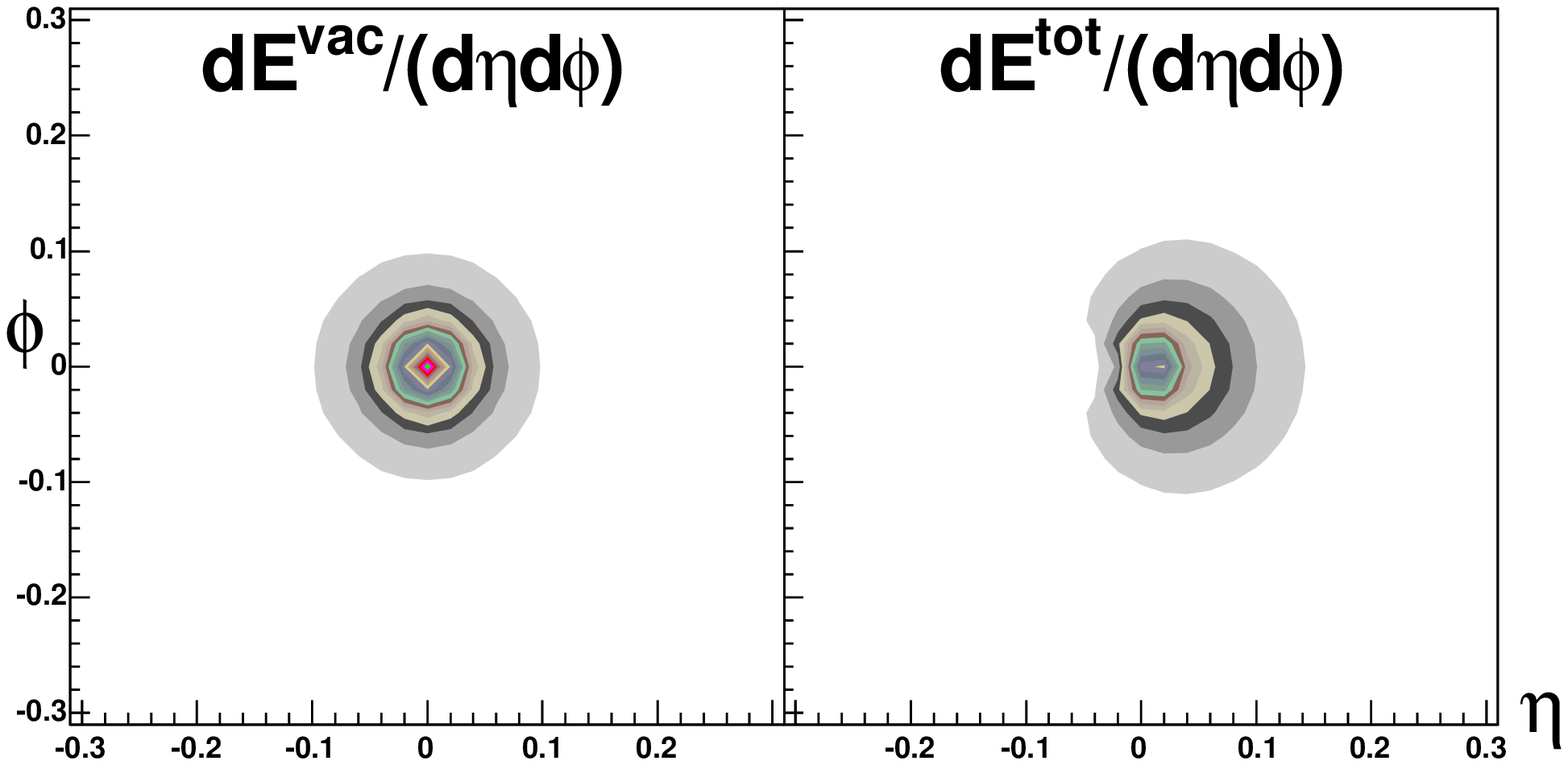}
\end{minipage}
\hfill
\begin{minipage}{0.46\textwidth}
\includegraphics[width=\textwidth]{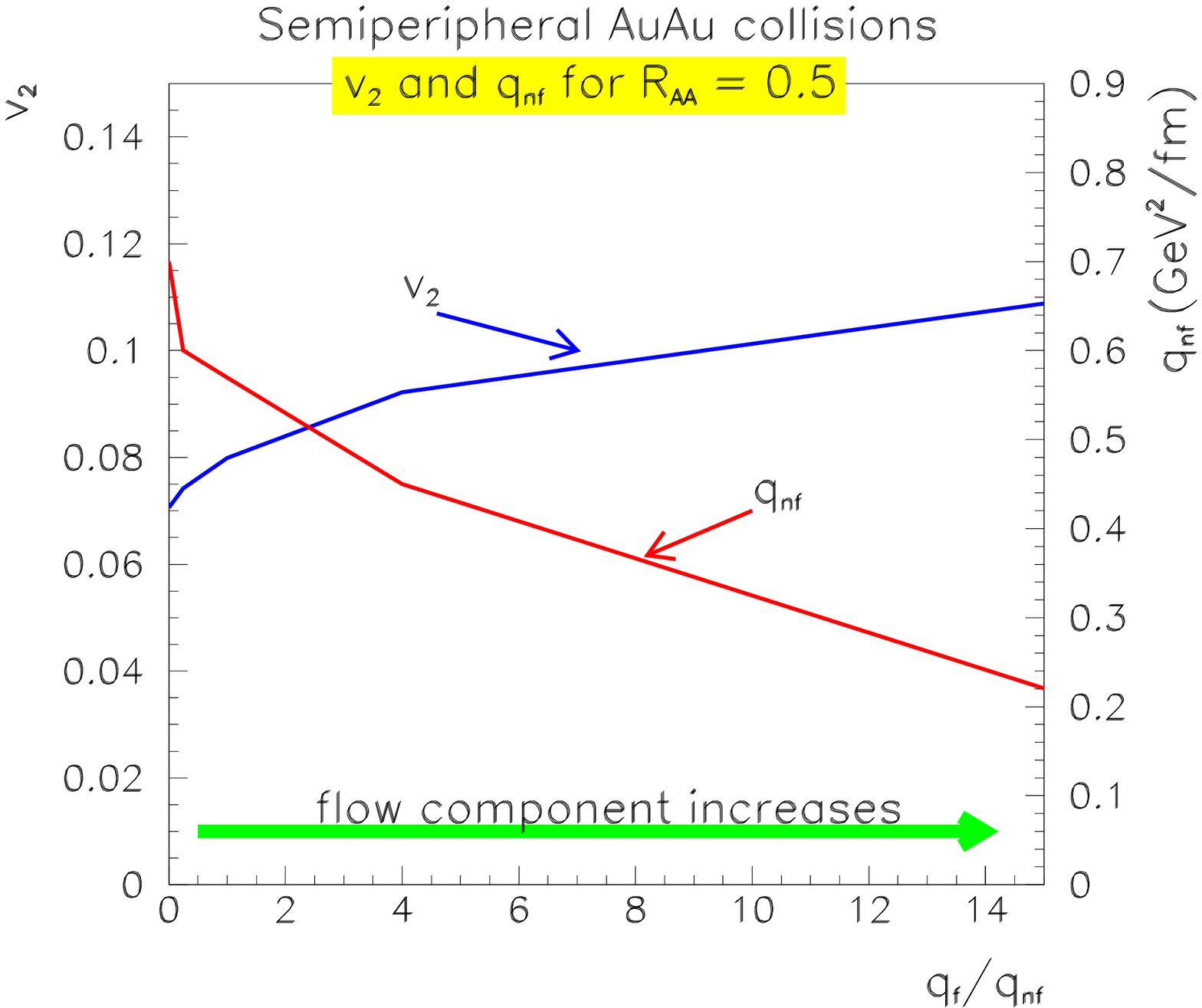}
\end{minipage}
\caption{Left: Jet energy distributions in the $\eta\times\phi$ plane for a
100 GeV
jet for a non-flowing (left panel) and flowing (right panel) medium. Right:
Dependence of $v_2$ and non-flow component for the same
$R_{AA}=0.5$ in semiperipheral AuAu collisions.}
\label{figflow}
\vskip -0.3cm
\end{figure}

At RHIC, a series of measurements on high-$p_t$ particle correlations 
\cite{qm05} point to
the interesting possibility that the parton shower is strongly modified by
dynamical medium properties as flow. From a theoretical point of view, the
fundamental quantity describing a medium is the energy-momentum tensor
$T^{\mu\nu}=(\epsilon+p)u^\mu u^\nu-p g^{\mu\nu}$ 
and this must, hence, be the object which determines $\hat q$. In a static
medium we should recover the proportionality $\hat q=c\,\epsilon^{3/4}$
\cite{Baier:2002tc}, with
$\epsilon$ the energy density. In the case of a transverse flow, the simplest
extrapolation gives $\hat q=c'\,(T^{n_\perp n_\perp})^{3/4}$, where now
$T^{n_\perp n_\perp}$ is given both by a non-flow symmetric
component (density) and a antisymmetric
flow field. One general conclusion is then that a flow field appears as an
additional source of medium-induced gluon radiation, which mimics the effect
of a static medium with larger density (see Fig. \ref{figflow}).
Moreover, the most clear signature in the case the flow field is strong
enough, is the distortion of the jet shapes
in the direction of the local flow field. Indeed, the associated radiation in
the presence of a flow is no longer azimuth-symmetric as seen in Fig.
\ref{figflow}. This opens new possibilities for studying dynamical properties 
of the produced medium by jet measurements.

\end{document}